\documentclass[twocolumn,showpacs,preprintnumbers,footinbib,prd,superscriptaddress,groupedaddress,10pt]{revtex4-1}

\usepackage[utf8]{inputenc}

\makeatletter
\def\l@subsubsection#1#2{}
\def\l@subsubsubsection#1#2{}
\makeatother

\usepackage{graphicx,amssymb,amsmath,amsthm,amsfonts}
\usepackage[usenames]{color}

\usepackage{bm}
\usepackage{dcolumn}
\usepackage{latexsym}
\usepackage{rotating}
\usepackage{longtable}

\usepackage{enumerate}
\usepackage{tensor,multirow}
\usepackage{url}
\usepackage[linktocpage]{hyperref}
\usepackage[export]{adjustbox}

\def\nn{\nonumber}

\def\be{\begin{equation}}
\def\ee{\end{equation}}
\def\beq{\begin{eqnarray}}
\def\eeq{\end{eqnarray}}

\begin{document}

\title{Geodesic structure and quasinormal modes of a tidally perturbed spacetime}

\author{Vitor Cardoso}
\affiliation{CENTRA, Departamento de F\'{\i}sica, Instituto Superior T\'ecnico -- IST, Universidade de Lisboa -- UL,
Avenida Rovisco Pais 1, 1049 Lisboa, Portugal}
\author{Arianna Foschi}
\affiliation{CENTRA, Departamento de F\'{\i}sica, Instituto Superior T\'ecnico -- IST, Universidade de Lisboa -- UL,
Avenida Rovisco Pais 1, 1049 Lisboa, Portugal}
\affiliation{Faculdade de Engenharia, Universidade do Porto, Rua Dr. Roberto Frias s/n, 4200-465 Porto, Portugal}

\begin{abstract} 
Tidal perturbations play an important role in the study of the dynamics in the classical two-body system. Understanding tidal effects in strong-field regions may allow one to use gravitational-wave or electromagnetic observations to locate or constraint the location of possible companions.
Here, we investigate how timelike and null geodesics of a Schwarzschild black hole are affected in the presence of a companion. There is a panoply of new effects.
In some limiting cases, we find analytical solutions for closed null or timelike geodesics. Our results show that light ring period as measured by a far-away observer can be eiter shorter or longer, depending on the location of the companion. We also show that there are closed lightlike trajectories which are elliptic (for equatorial companions), and that timelike particles are affected in a similar manner.
Finally, we attempt at estimating the ringdown from tidally perturbed geometries. Our results indicate that there are two stages
in the relaxation of such geometries, one associated with a prompt decay of waves around the deformed photonsphere, and a later 
relaxation of the global geometry. These results are consistent with previous, full numerical studies.
\end{abstract}

\maketitle

\tableofcontents
\section{Introduction}
Tidal interactions play a fundamental role in many astrophysical systems. The best known example are the oceans tides in the Earth-Moon system, which drive tidal acceleration and longer days,
but their effects are ubiquitous in astrophysics. In the context of gravitational-wave (GW) astronomy, of tests of fundamental physics and of strong-field gravity, tidal effects are precious.
In a compact binary emitting GWs, tidal deformations induced by the companion affect the GW phase. The amount of the de-phasing (relative to that of pointlike objects) correlates to the equation of state of the inspiralling bodies, hence a precise monitoring of the GW phase evolution can teach us about the equation of state of neutron star binaries~\citep{Flanagan:2007ix,Dietrich:2020eud,Gupta:2020lnv}. Black holes (BHs) have a particularly simple equation of state, owing to the uniqueness properties in vacuum General Relativity~\cite{Chrusciel:2012jk,Cardoso:2016ryw}. Hence, the tidal interactions of compact objects are particularly useful to test the Kerr nature of BHs~\cite{Cardoso:2017cfl,Sennett:2017etc,Cardoso:2019rvt,Cardoso:2019rvt}.

Previous studies explored the potential of tidal interactions to constrain the presence of a possible massive companion to the Sgr*A source~\cite{Naoz:2019sjx}.
Specifically, data collected by the GRAVITY collaboration~\cite{Abuter:2018drb,Abuter:2020dou} on the orbital motion of the star S0-2 was used to constrain possible orbital parameters of such a companion. It was also shown that a putative companion may give rise to GW emission potentially detectable with the future space-based interferometer LISA.
The effect of tidal fields in Newtonian orbits is, of course, well studied, specially in the restricted three-body problem (see Refs.~\cite{murray_dermott_2000,Domingos:2008} and references therein).
Here, we want to understand the effects of weak tides in the strong-field region. We are particularly interested on the effect of tides on the location and properties of the innermost stable circular orbit (ISCO), and of the photonsphere. These dictate the high-energy behavior of accretion disks, and the relaxation properties of BHs~\cite{MTB,Cardoso:2008bp,Cardoso:2016rao,Yang:2021zqy}, and so can be useful smoking-guns of companions. Driven by similar motivations, this problem was studied recently, albeit in less realistic setups. The position of the ISCO was studied in a spacetime describing a binary of extremal charged BHs, in particular the Majumdar-Papapetrou dihole spacetime~\cite{Nakashi:2019mvs,Nakashi:2019tbz}. The photonsphere of binary BH spacetimes has been the subject of recent studies~\cite{Assumpcao:2018bka,Bernard:2019nkv,Daza:2018nsk}, since it may help in understanding the relaxation or ringdown stage of binaries themselves, or hold the key to understanding how the individual components quasinormal ringdown is affected by the companion. These spacetimes include, naturally, tidal effects, but their effects have mostly been studied numerically. Our aim is to provide simple analytical results of strong field phenomena. We use geometric units $G=c=1$ throughout.

\section{Setup: a black hole perturbed by a companion}\label{sec_setup}
Consider a non-spinning BH of mass $M$, perturbed by a companion of mass $M_c$ at a distance $R$. For far-away companions, the tide is weak and one can expand the geometry around its Schwarzschild value, 
\begin{equation}
g_{\mu \nu} = g_{\mu \nu}^{\rm Sch} + \epsilon \, h_{\mu \nu}\,,\label{full_metric}
\end{equation}
where $\epsilon$ represents the strength of the tidal perturbation, assumed to be small ($\epsilon \ll 1$). 
Using Regge and Wheeler's approach, the perturbation $h_{\mu \nu}$ can be expanded in tensor spherical harmonics~\cite{Regge:1957td,Zerilli:1971wd,Berti:2009kk}
%
%
and decomposed in polar and axial parts, due to the spherical symmetry of the system. We focus on the polar contributions which, in the Regge-Wheeler gauge, can be written as:
%
\begin{eqnarray}
h_{\mu \nu}^{\rm polar} = \left( \begin{matrix} f  H_0^{l m} & H_1^{l m}\, & 0 & 0 \\
H_1^{l m}  & f^{-1} H_2^{l m} & 0 & 0 \\
0 & 0 & r^2  K^{l m}  & 0 \\
0 & 0 & 0 & r^2 \sin^2 \theta K^{l m} \end{matrix} \right) Y^{l m} \,\,\label{hpolar}
\end{eqnarray}
where $H_i^{l m}=H_i^{l m}(t,r)$, $K^{l m}=K^{l m}(t,r)$, $Y^{lm}=Y^{l m}(\theta, \phi)$ are standard spin-0 spherical harmonics and 
\begin{equation}
f=f(r) =\left(1-\frac{2M}{r}\right)\,.
\end{equation}
%
%

To continue, consider a companion far away, such that the orbital timescale is larger than any other scale in the problem and much larger than the timescale of the internal dynamics of the body.
In this regime, one can focus on static perturbations. 
The vacuum field equations then provide equations for the metric functions $H_0, H_1, H_2, K$.
These should be solved demanding regularity at the horizon. We will mostly focus on quadrupolar modes, but the analysis is easily extended.
The $tr$ and $\theta \theta$ components of the field equations result in
\be
H_1=0\,,\qquad H_2=H_0\,.
\ee
The solution for $H_2$ which is regular across the horizon $r=2M$ for $l=2$ is given by
\beq
H_2 =\, c_1 \frac{3\, (2 M - r)\, r }{M^2} = - \frac{3 \, c_1}{M^2}\, r^2 \, f(r)\,,\label{h2}
\eeq
where $c_1$ is an integration constant that must be determined by studying the asymptotic behaviour of the metric \eqref{full_metric} in the presence of an external tidal field.  
This can be done using the definition of multipole moments developed by Thorne~\cite{Thorne:1980ru,Zhang:1986cpa}. The effects of an external tidal field are entirely encoded in two symmetric and trace-free tensors: the polar tidal field $\mathcal{E}_L$ and the axial tidal field $\mathcal{B}_L$. The polar tidal field can be expanded in spherical harmonics as
$\mathcal{E}_L x^L =r^l \sum_m \mathcal{E}_{lm} Y^{lm}(\theta_c, \phi_c)$, where $\theta=\theta_c$ and $\phi=\phi_c$ are the angular coordinates of the companion in the BH sky~\cite{Binnington:2009bb}.

In the most general case,  the asymptotic expansion of the metric will depend on the angular index $m$.  Since we only need it to find the value of $c_1$, we can fix $m=0$ without loss of generality and the metric expansion reads
%
\begin{equation}
\begin{split}
g_{tt} &= -1 + \frac{2 M}{r} + \sum_{l \geq 2} \left[ \frac{2}{r^{l+1}} \left(\sqrt{\frac{4 \pi}{2l + 1}} M_l Y^{l 0} + (l'< l \, \text{pole}) \right) \right. \\
& - \left. \frac{2}{l(l-1)} r^{l} [ \mathcal{E}_{l} Y^{l0} + (l' < l\, \text{pole})] \right]\,, \label{gttasym}
\end{split}
\end{equation}
%
where $M_l$ are the mass multipole moments.

If we consider the dominant quadrupolar contribution with $l=2$, we can match Eq.~\eqref{gttasym} with the $tt$-component of Eq.~\eqref{full_metric}, and find
\begin{equation}
c_1=\frac{M^2 \, \mathcal{E}_2}{3} \, .
\end{equation}
Now we have to find the explicit value of the tidal moment $\mathcal{E}_2$.   
This can be done matching the full metric ~\eqref{full_metric} with a Post Newtonian (PN) description of the external spacetime \cite{Binnington:2009bb, PoissonWill}.   
We will always assume parameters for which the system can be captured by a PN description.
The $g_{tt}$ component in the PN approximation is~\cite{PoissonWill}:
\begin{equation}
g_{tt}^{\rm PN}=-1+2U(r,\theta,\phi)\,.\label{gttpn}
\end{equation}
If we assume that the companion is a PN monopole of mass $M_c$ and centering ouselves in the BH frame, the potential can be written as
\begin{eqnarray}
U &=& \frac{M}{r} + \frac{M_c}{|\mathbf{r}- \mathbf{R}|} \nn \\  
&=&  \frac{M}{r} +  M_c \sum_{lm} \frac{4 \pi}{2l +1} \frac{r^l}{R^{l+1}}Y^*_{lm}(\theta_c, \phi_c) Y_{lm} (\theta, \phi).\label{laplace_potential}
\end{eqnarray}
Finally for $l=2$ we have
\begin{equation}
g_{tt}^{\rm PN}=-1+\frac{2M}{r}+\frac{8 \pi M_c}{5}\frac{r^2}{R^3}\,\sum_m Y^*_{2 m}(\theta_c, \phi_c)\, Y_{2 m}(\theta, \phi) \,,\label{gttpnfinal}
\end{equation}
and comparing~\eqref{gttpnfinal} with the $g_{tt}$ component of the perturbed metric we find that
\begin{equation}
\mathcal{E}_2 = - \frac{8 \pi  M_c }{5 R^3} \sum_m \, Y^*_{2 m}(\theta_c, \phi_c) = - \frac{8 \pi \epsilon }{5 M^2}\, \sum_m Y^*_{2 m}(\theta_c, \phi_c)\,,\label{epsilon2}
\end{equation}
where in the last step we have defined the strength of the tidal deformation as 
\be
\epsilon = M^2 M_c/R^3\,.\label{epsilon}
\ee
Equation~\eqref{epsilon2} agrees with the expression for $\mathcal{E}_2$ in \cite{Cardoso:2020hca}. The metric functions in~\eqref{hpolar} are now completely determined and they read:
\beq
H_0 (r) &=& H_2 (r) = \frac{8 \pi \epsilon }{5 M^2}  \, r^2 f(r) \sum_m Y^*_{2 m}(\theta_c, \phi_c)\,,  \\
K(r) &=& \frac{8 \pi \epsilon }{5 M^2}  \left(r^2-2M^2\right) \sum_m Y^*_{2 m}(\theta_c, \phi_c) \, .
\eeq
%
\section{Polar companions}\label{sec_polar}
We will specialize our calculations to two specific setups, where the orbits lie
in the BH-companion plane, or orthogonal to it. Equivalently, and this is the approach we follow,
we restrict to equatorial orbits, and place the companion either at the equator or at the pole.

We start with a companion at the pole, $\theta_c = 0, \, \phi_c =0$, which then preserves the azimuthal symmetry of the BH.
\subsection{ISCO and light ring properties}
The full components of the metric with $l=2$ on the equatorial plane are:
\beq
g_{tt}^{\rm polar} &=& -f(r) \left[ 1 + f(r) \, \frac{r^2 \, \epsilon}{M^2} \right]\,,\label{gtt_pert}\\
g_{rr}^{\rm polar} &=& g(r) -\frac{ r^2 \, \epsilon}{M^2}\,, \label{grr_pert}\\
g_{\theta \theta}^{\rm polar} &=& g_{\phi \phi}^{\rm polar} = r^2 \left[1 -\frac{ (r^2 -2M^2) }{M^2}\, \epsilon \right] \label{gfifi_pert}\,.
\eeq
In the equatorial plane, we then find the Lagrangian
\begin{equation}
2\mathcal{L}=g_{tt}^{\rm polar} \dot{t}^2+ g_{rr}^{\rm polar} \dot{r}^2 + g_{\phi \phi}^{\rm polar}\dot{\phi}^2\,,
\end{equation}
with dots standing for derivatives with respect to proper time.
The Lagrangian is time and azimuth independent, giving rise to two conserved quantities, specific energy $E$ and angular momentum $L$
\beq
\dot{t} &=&\frac{E}{f(r)(1+\epsilon f r^2/M^2 )} \,, \label{tdot} \\
\dot{\phi} &=&\frac{L}{r^2(1+(2-r^2/M^2)\, \epsilon)} \, .\label{phidot}
\eeq

The equations of motion are easier to handle via an effective radial potential, which can be obtained substituting \eqref{tdot}-\eqref{phidot} in the normalization for the quadrivelocity
\be
g_{\mu \nu} \, u^{\mu} u^{\nu} = \delta\,,
\label{norm_quadr}
\ee
where $\delta = 0,-1$ for null and timelike geodesics, respectively. 
One finds,
\be
\dot{r}^2=E^2-V_{\delta}(r)\,,\label{rdot}
\ee
where
\beq
V_{\delta}(r)&=&\frac{L^2 \, (2M-r)\left( 2 M r \epsilon -2 r^2 \epsilon + M^2(2 \epsilon -1) \right)}{M^2 r^3} \nonumber\\
&+& \delta \frac{(2M-r)\, (M^2 -2 M r \epsilon + r^2 \epsilon )}{M^2 r} \,.\label{potential}
\eeq

To understand the ISCO properties, we expand all relevant quantities in Eq.~\eqref{potential} to first order in $\epsilon$, e.g.,
$r_{\rm ISCO} = r^{(0)} + \epsilon \, r^{(1)}$, with $r^{(0)}=6M$ the unperturbed Schwarzschild BH value.
By definition, at the ISCO $E^2-V_{(\delta=-1)}=d V_{(\delta=-1)}/dr= d^2 V_{(\delta=-1)}/dr^2= 0$. 
We find
\beq
r_{\rm ISCO}&=& 6M(1-256\epsilon) \,, \label{ISCO}\\
E_{\rm ISCO}&=& \frac{2 \sqrt{2}}{3}\left(1+38\epsilon\right) \,,\label{E_ISCO}\\
L_{\rm ISCO}&=& 2 \sqrt{3}M(1+ 7\epsilon) \,,\label{L_ISCO}\\
\Omega_{\rm ISCO}&=& \frac{1+491\epsilon}{6 \sqrt{6}\, M} \,. \label{omega_ISCO}
\eeq
Note that $\Omega_{\rm ISCO}$ is simply $\dot{\phi}/\dot{t}$ evaluated at the ISCO, and is the angular velocity as measured by far away observers.

\begin{table}[!htbp]
\caption{\label{table_comparison} 
Numerical results for the ISCO properties of a tidally deformed BH spacetime. The analytical results \eqref{ISCO}-\eqref{omega_ISCO} agree with these values up to the last digit.
}
\begin{ruledtabular}
\begin{tabular}{cccc}
\multicolumn{1}{c}{}           & \multicolumn{3}{c}{\textbf{Numerical results}} \\ \hline
\multicolumn{1}{c}{$\epsilon$} & $r/M$   & $E$      & $L/M$      \\ \hline 
\multicolumn{1}{c} {$10^{-7}$} & 5.9998  & 0.94281  & 3.4641   \\ 
\multicolumn{1}{c}{$10^{-6}$}  & 5.9985  & 0.94284  & 3.4644       \\ 
\multicolumn{1}{c}{$10^{-5}$}  & 5.9849  & 0.94317  & 3.4671     \\ 
\multicolumn{1}{c}{$10^{-4}$}  & 5.8657  & 0.94633  & 3.4932   \\ 
\end{tabular}
\end{ruledtabular}
\end{table}
High frequency photons or gravitons are well described by null geodesics. Of these, there is one that stands out: a close null geodesic
which for non-rotating, isolated BHs is located at $r=3M$~\cite{MTB,Cardoso:2008bp,Cardoso:2016rao,Cardoso:2019rvt,Yang:2021zqy}.
This location defines the light ring or photosphere. In the presence of a companion, we find
\beq
r_{\rm LR}&=& 3M(1+ 5\epsilon) \,,\label{light_ring}\\
b_{\rm LR}&=& 3 \sqrt{3}M(1-5 \epsilon)M\,,\label{b_LR}\\
\Omega_{\rm LR}&=&\frac{1+ 5 \, \epsilon}{3 \sqrt{3} \, M} \,. \label{omega_LR}
\eeq
where $b = L/E$ is the impact parameter.

These analytical estimates can be compared to a numerical solution of the geodesic equations. Those results are shown in Table~\ref{table_comparison} for some selected values of $\epsilon$.
The perturbative analytical results agree with these numbers to all digits listed. These results can be compared and contrasted to those referring to an extremally-charged BH binary, the Majumdar-Papapetrou geometry. This is done in Appendix~\ref{app:MP}.

\subsection{The relaxation of tidally perturbed black holes: light ring modes and quasinormal modes}
Consider now fundamental fields in the BH vicinities, such as gravitational or electromagnetic waves.
The dynamical evolution is described by a set of second order differential equations whose details depend
on the initial data. In general however, the time evolution is characterized by a prompt signal, followed by a ringdown
which is caused by the ``leaky'' boundary conditions at the boundaries~\cite{Leaver:1986gd,Berti:2009kk}.
When new structure is added, such as a change of boundary conditions at large distances or in the near-horizon region, new features appear. In particular,
the light ring controls the early-time relaxation of BHs, whereas the late-time ringdown is dictated
by boundary conditions, defining quasinormal modes~\cite{Cardoso:2008bp,Yang:2012he,Cardoso:2016rao,Cardoso:2017cqb,Cardoso:2019rvt}.
For isolated BHs, these two decays coincide and the dynamical properties of BHs are relatively simple.

We are dealing with a tidally perturbed BH, so we expect two stages in the dynamical evolution of fundamental massless fields. In fact, there is evidence in the literature 
for such distinctive behavior~\cite{Bernard:2019nkv}.
Consider, first, the light ring relaxation. A complete knowledge must be obtained performing time evolutions of the corresponding evolution equations, but a good description
is obtained in the eikonal approximation~\cite{Cardoso:2008bp,Yang:2012he}. In this approximation, the early time ringdown of a signal $\Phi$ is an exponentially damped sinusoid,
\be
\Phi\sim e^{-\omega_I^{\rm LR} t}\sin{\omega_R^{\rm LR}t}\,.
\ee
Formally, the signal is a superposition of overtones which carry an associated index $\ell$ associated with the angular dependence and an index $n$ associated to a radial dependence.
Here, the ringdown frequency $\omega_R^{\rm LR}$ and the damping rate $\omega_I^{\rm LR}$ are given by the angular frequency \eqref{omega_LR} and the Lyapunov exponent $\lambda=\sqrt{V_r^{''}/2 \, \dot{t}^2}$, respectively~\cite{Cardoso:2008bp}.
 
 Although this limit is formally valid only when $\ell \gg 1$, it has been shown that it gives very accurate results even when $\ell$ is smaller~\cite{Berti:2009kk}. In particular for modes with $\ell=m$ we have $\omega_R = \ell \, \Omega_{\rm LR}$ and $\omega_I =(n + 1/2) |\lambda_{\rm LR}|$. For $\ell=m=2$ the ringdown frequency and the damping rate are
\be
M \omega_R= \frac{2 \, (1 + 5\, \epsilon)}{3 \sqrt{3}}\,,\quad M \omega_I = \left(n +\frac{1}{2} \right) \frac{1 - 10\, \epsilon}{3\, \sqrt{3}} \,.\label{LR_ringdown}
\ee
We thus have a clear prediction for the changes in the early ringdown of GWs, induced by a companion.

\begin{figure*}[ht!]
 \centering
 \includegraphics[width=1\textwidth, valign=T]{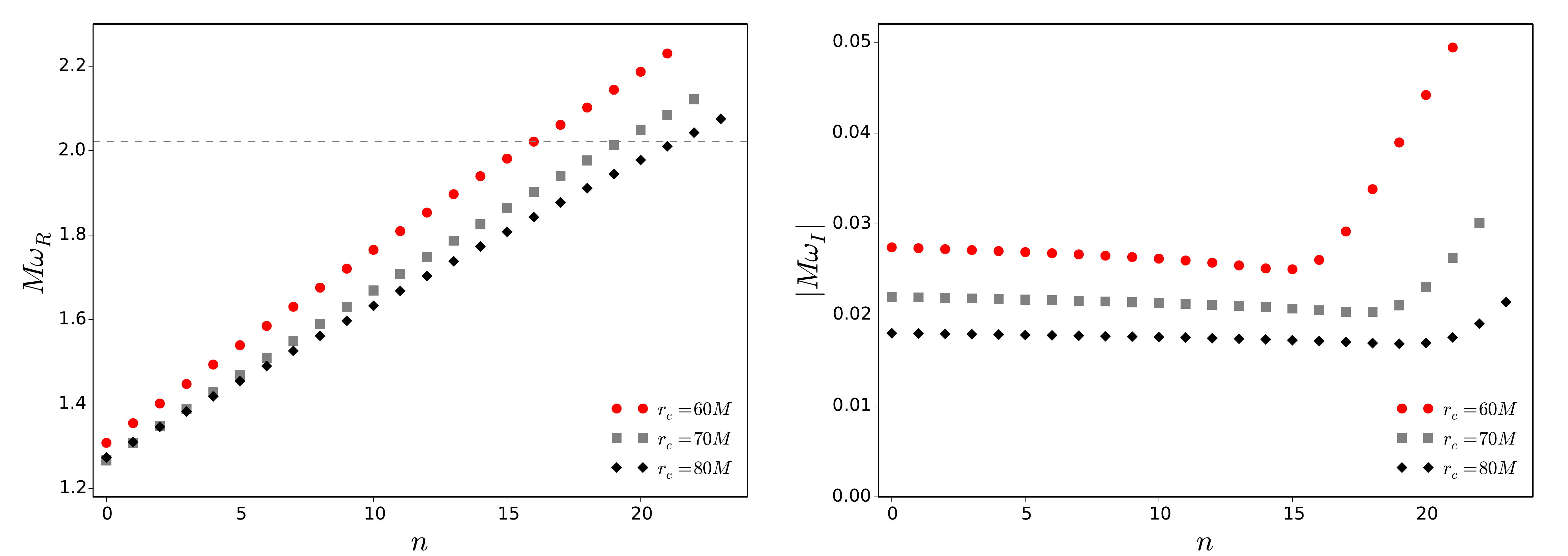}
\caption{Real (left panel) and imaginary (right panel) part of the frequencies for different values of the ``cutoff point'' $r_c$ in the function $H(r)$. In this case $\epsilon = 10^{-4}$, $k=100$, and $l=m=10$. The dashed line represents the Schwarzschild frequency (not shown in the right panel since its value $|M \omega_I|^{\rm Sch} \simeq 0.09$ is out of the range chosen).}
\label{Fig:freqsk100}
\end{figure*}
\begin{figure*}[ht!]
 \centering
 \includegraphics[width=1\textwidth, valign=T]{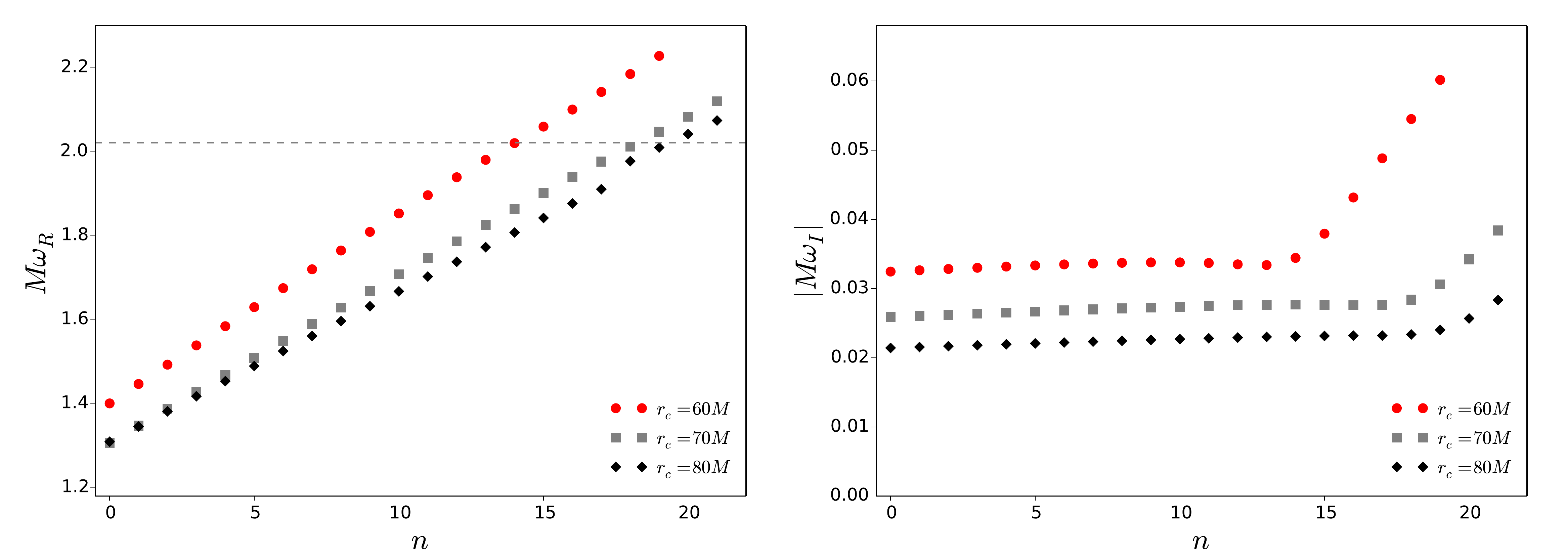}
\caption{Real (left panel) and imaginary (right panel) part of the frequencies for different values of the ``cutoff point'' $r_c$ in the function $H(r)$. In this case $\epsilon = 10^{-4}$, $k=2$ and $l=m=10$. The dashed line represents the Schwarzschild frequency (not shown in the right panel since its value $|M \omega_I|^{\rm Sch} \simeq 0.09$ is out of the range chosen). }
\label{Fig:freqsk2}
\end{figure*}
Now, as we mentioned, the late time behavior of the GW signal is dominated by the poles of the relevant Green function the quasinormal modes (QNMs), and therefore
are sensitive to the entire geometry and not only to the local properties around the light ring.
In order to assess the QNM spectrum one needs to solve the dynamics. To understand the possible changes, we focus on the relatively simpler
problem of a scalar field propagating on the fixed, background geometry~\eqref{full_metric}.
The dynamics are governed by the Klein-Gordon equation,
\beq
\Box \varphi = \frac{1}{\sqrt{-g}} \partial_{\mu} \left(g^{\mu \nu}_{\rm polar} \sqrt{-g} \, \partial_{\nu} \varphi \right) = 0 \,,\label{KG_eq}
\eeq
where $g = \det{g^{\rm polar}_{\mu \nu}}$. Since the spacetime admits two Killing vectors $\partial_t$ and $\partial_{\phi}$, we can decompose the scalar field as:
\beq
\varphi(t, r, \theta, \phi) = e^{i (m \phi -\omega t)} \frac{\psi_l(r)}{r} Y_{lm}(\theta) \,.
\label{scalar_field}
\eeq
If we substitute Eq.~\eqref{scalar_field} in Eq.~\eqref{KG_eq} we get a system of differential equations that are not separable. 
Following Ref.~\cite{Cano:2020cao}, we can expand the box operator in \eqref{KG_eq} at first order in $\epsilon$:
\beq
\Box = \Box_{(0)}  + \epsilon \, \Box_{(1)}\,,\label{box}
\eeq 
where we have taken into account that the zeroth order part is separable while the non-separable terms come only from the perturbative correction to the metric. The explicit form of the operator is
\beq
&&\Box_{(0)} \varphi_{m, \omega}(r, \theta) = \frac{1}{r^2} \partial_r \left( r(r-2M) \partial_r \varphi \right)  \nonumber \\
&+& \frac{1}{r^2  \sin \theta} \partial_{\theta} \left( \sin \theta \, \partial_{\theta} \varphi \right) + \left[ \frac{r \, \omega^2}{r-2M} - \frac{m^2}{r^2 \sin^2 \theta}\right] \varphi\,,
\eeq
\beq 
&&\frac{2M^2}{1+3 \cos 2\theta}\Box_{(1)}\varphi_{m, \omega}(r, \theta) =\frac{2M^2-r^2}{r^2} \cot\theta \partial_{\theta} \varphi \nonumber \\
&+& \frac{2M^2-r^2}{r^2} \partial_{\theta}^2 \varphi-\frac{2(r-2M)^2}{r} \partial_r \varphi - (r-2M)^2 \partial_r^2 \varphi \nonumber \\
&+& \left(r^2 \omega^2 + \frac{(r^2 -2M^2)}{r^2 \sin^2 \theta} m^2\right)\varphi\,.
\eeq
Since the zeroth-order solution for a given $l_0$ is $\varphi(r, \theta) = \psi_{l_0}(r)/r \, Y_{l_0 m}(\theta)$, it is possible and consistent to assume that any term with $l \neq l_0$ comes from the perturbative correction. Hence, we can expand the solution such as,
\beq
\varphi_{m, \omega}(r, \theta) = \frac{\psi_{l_0}(r)}{r} Y_{l_0 m}(\theta) + \epsilon \sum_{l \neq l_0}   \frac{\psi_l(r)}{r} Y_{lm}(\theta)\label{expanded_solution}
\eeq
Applying the operator (\ref{box}) to the solution (\ref{expanded_solution}) and negleting terms $\mathcal{O}(\epsilon^2)$ we have:
\beq
& & \Box_{(0)} \left[ \frac{\psi_{l_0}(r)}{r} Y_{l_0 m}(\theta)\right] + \epsilon \sum_{l \neq l_0} \Box_{(0)} \left[ Y_{lm}(\theta)  \frac{\psi_l(r)}{r} \right] +  \nonumber \\
& & + \epsilon \, \Box_{(1)} \left[ \frac{\psi_{l_0}(r)}{r} Y_{l_0 m}(\theta) \right] + \mathcal{O}(\epsilon^2) = 0 \,.
\label{KG_expanded}
\eeq
Projecting (\ref{KG_expanded}) onto $Y_{l_0 m} (\theta)$ we get one single radial equation that can be solved numerically to find the QNM frequencies:
\beq
& & \left(\Box_{(0)} - l(l+1) \right) \frac{\psi_{l_0}(r)}{r}  \nonumber \\
 &+& \epsilon \int_{0}^{\pi} \Box_{(1)} \left[\frac{\psi_{l_0}(r)}{r} Y_{l_0 m}(\theta)\right] Y_{l_0 m} (\theta) d \theta = 0 \, .
\label{KG_final}
\eeq

The asymptotic behavior of the solutions of the above equation are as follows, 
\beq
\psi_{l_0}&\sim&(r-2M)^{2M i \omega}\,,\qquad r\to 2M\,,\\
\psi_{l_0}&\sim&\,e^{\pm \omega r}\,, \qquad r\to \infty\,.
\eeq
The boundary condition at the horizon is similar to that of isolated BHs~\cite{Berti:2009kk}. 
However, the boundary behavior at large spatial distances, imposed on us by the equations of motion, are completely different from those of an asymptotically flat spacetime. 
Indeed, the metric perturbation of the tidally perturbed BH we're studying is {\it not} asymptotically flat. We are using the diverging piece of the metric perturbation
to impose the presence of a companion far away. In a complete setup, the companion is at finite distance and the spacetime asymptotically flat.
In other words, our description of the binary system is only accurate for distances $r \lesssim R$.   
The boundary conditions imposed on us within this setup are similar to those of asymptotically anti-de Sitter or other spacetimes where radiation is confined~\cite{Berti:2009kk,Brito:2014nja}.
They select a set of complex quasinormal mode frequencies
\be
\omega=\omega_{R}+i\omega_{I}\,.
\ee
Accordingly, the confining nature of the tidal perturbations indicates that one will find a QNM spectrum which differs substantially from that of a single Schwarzschild BH, reflecting the fact that perturbations should be less damped
(as the only dissipation channel is now the horizon).

We thus need to fix the unacceptable behavior at large spatial distances. We will do so without entering the challenge of matched asymptotic expansions to correctly reproduce
the geometry everywhere~\cite{Bernard:2019nkv}. We simply ``cutoff'' the tidal effects with an auxiliary function, which we take rather arbitrarily to have the form
$H(r) = 1/1+e^{2 k (r- r_c)}$. In other words, our regularization procedure consists on the replacement
\be
\epsilon \rightarrow  \frac{\epsilon}{1+e^{2 k (r- r_c)}}\,,
\label{cutoff_function}
\ee
since the $\epsilon$ terms are precisely the ones responsible for the divergence at infinity. 
In general, results will depend on the smoothness parameter $k$ and the \textit{cutoff point} $r_c$. It seems reasonable to ask that $r_c\sim R$, whereas $k$ is at least of the order $M_c$ to describe
a smooth transition.

Figures~\ref{Fig:freqsk100}-\ref{Fig:freqsk2} show the QNMs for three different values of the cutoff radius $r_c/M=60,\, 70,\, 80$ and $k=100$ (top panels) or $k=2$ (bottom panels).
Notice that what is shown are the first overtones.
Since we are mostly interested in comparing against the eikonal limit, we fix $\ell=m=10$. 
We fixed a tidal parameter $\epsilon = 10^{-4}$, but results for other $\epsilon$ are similar.

As already anticipated from the discussion above, the most salient feature of Figs.~\ref{Fig:freqsk100}-\ref{Fig:freqsk2} is that the QNM spectrum of a tidally deformed BH is completely different from that of an isolated BH.  Despite our curing the asymptotic behavior artificially, remains of this behavior remain in the perturbation via the existence of quasi-bound states.
For example, for a perturbation parameter $\epsilon=10^{-4}$, one could expect a correspondingly small change in the QNM spectrum. However, as can be seen in Figs.~\ref{Fig:freqsk100}-\ref{Fig:freqsk2}, the QNM frequencies change by ${\cal O}(1)$.
These features were seen in the past~\cite{Leung:1999rh,Barausse:2014tra,Brito:2014nja,Cardoso:2016rao,Cardoso:2019rvt} and are connected with the asymptotic properties of the effective potential for wave propagation
(and of the corresponding solutions). In fact, the tidal effects act to create a long-distance ``well'' that traps low frequency radiation. This explains why the damping of the lowest modes (see right panels in Figs.~\ref{Fig:freqsk100}-\ref{Fig:freqsk2}) is so much lower than the isolated-BH counterpart. Once the vibration frequency $\omega_R$ is sufficiently large for fluctuations to tunnel out through the light ring, $\omega_R \simeq \omega_R^{\rm Sch}$~\cite{Cardoso:2008bp}, a new channel for dissipation is open, and the damping timescale decreases.

Our results are well described by a real component
\beq
\omega_R \approx \omega_{R}^{(0)} + \alpha \,  n \,,\quad n=0,1,2...
\eeq
where the offset $\omega_R^{(0)}$ corresponds to the fundamental mode $n=0$, and $\alpha$ is a constant, which decreases increasing $r_c$ and it is independent (or only weakly dependent) on $\epsilon$. Our results indicate a very weak dependence of $\omega_{R}^{(0)}$ on $r_c$ and $\epsilon$. At very small $\epsilon$ we recover the Schwarzschild fundamental QNM, but it is not necessarily the fundamental mode of this perturbed spacetime. This is the spectrum characteristic of a confined system and differs markedly from that of an isolated BH, for which the (real part of the) QNM frequencies asymptote to a constant~\cite{Berti:2009kk}.
For $k=100$, $\alpha = 0.044, \, 0.039, \, 0.035$ for $r_c/M = 60,\,70,\, 80$ respectively, 
while for $k=2$, $\alpha = 0.052, 0.048, 0.041$.

Regarding the imaginary part of the QNM frequencies, our results for the fundamental mode are well described by
\be
M \omega_I^{(0)} \approx \frac{3 M}{5 \, r_c \, \epsilon^{1/8}} \,.
\ee
As we stressed, these results are sensitive to the cutoff function and radius we choose. We have also investigated the auxiliary function $H(r) = \Theta(r_c -r)$, with $\Theta$ the Heaviside function. Our results for the QNM spectrum show the same qualitative behavior as in Figs.~\ref{Fig:freqsk100}-\ref{Fig:freqsk2}. Although the numerical values are slightly different,
the overall structure is the same, and we find frequencies below the Schwarzschild fundamental mode. We also find a transition in the spectrum at $\omega_R \simeq \omega_R^{\rm Sch}$, as before. This lends strength to the claim that Figs.~\ref{Fig:freqsk100}-\ref{Fig:freqsk2} are a good qualitative description of the spectrum.

In conclusion, there are two different stages to consider in the relaxation of binaries~\cite{Bernard:2019nkv}. The first one is associated with a prompt ringdown phase and the local properties of the BHs light ring, which are affected by the presence of the companion via~\eqref{LR_ringdown}. This is a local property of the spacetime and hence is not affected by the asymptotic structure at infinity.  The second one is associated with the late time decay of the field which instead strongly depends on the global properties of the spacetime and that we tried to capture solving~\eqref{KG_final}. Although it is impossible with our naive cutoff procedure to fully capture all the physics, the structure of the spectrum confirms that we're dealing with a very different late-time relaxation
induced by the different scales in the problem. These features have already been seen in the numerical studies of Ref.~\cite{Bernard:2019nkv}.

\section{Equatorial companions}
\label{sec_equator}
We now specialize our analysis to equatorial companions, for which $\theta_c = \pi/2$. Without loss of generality we can impose $ \phi_c =0$. The presence of the companion, which is still assumed to be at rest, breaks the axisymmetry of the metric. As a consequence, the orbits that now lie on the BH-companion plane, have an explicit dependence on the azimuthal angle. For $\ell=2$, the metric components read:
%
\beq
g_{tt}^{\rm eq} &=& \frac{f(r)}{2 M^2}  \left[ - 2M^2 +f(r)\, r^2 \epsilon \left(1+3 \cos 2 \phi\right)\right]\,,\label{gtt_equat}\\
g_{rr}^{\rm eq} &=& \frac{g(r) }{2 M^2} \left[2M^2 - 2 M r \epsilon + r^2 \epsilon + 3 r (r-2M)\epsilon \cos2\phi\right] \,, \label{grr_equat}\\
g_{\theta \theta}^{\rm eq} &=&  g_{\phi \phi}^{\rm eq}\nonumber\\
& =& \frac{r^2}{2 M^2}  \left[2M^2(1-\epsilon) + r^2 \epsilon + 3(r^2-2M^2) \epsilon \cos 2 \phi \right] \label{gfifi_equat}\,.
\eeq

\subsection{Null geodesics}
%
\begin{table}[!htbp]
\caption{\label{table_lightring} Coordinate value of $r_0$ at $\theta=\pi/2, \,\phi=0$ for which the null geodesic is closed. We show both the numerical result $r_0^{\rm num}$
and the analytical prediction $r_0^{\rm an}$ given in Eq.~\eqref{r_phi_true} and valid for small $\epsilon$.}
\begin{ruledtabular}
\begin{tabular}{ccc}
$\epsilon$ &$3- r_0^{\rm num}/M$ & $3-r_0^{\rm an}/M$ \\ \hline 
$10^{-6}$ &$1.2\times 10^{-5}$ & $1.2\times 10^{-5}$  \\
$10^{-5}$ &$1.2\times 10^{-4}$ & $1.2\times 10^{-4}$ \\ 
$10^{-4}$ &$1.2\times 10^{-3}$ & $1.2\times 10^{-3}$\\  
$10^{-3}$ &$1.2\times 10^{-2}$ & $1.2\times 10^{-2}$ \\
\end{tabular}
\end{ruledtabular}
\end{table}
Due to the non-axisymmetry of the spacetime, solving geodesic motion analytically is challenging. Instead, we perform a numerical integration of geodesic equations and find the best fit for the motion's parameters. Consider first null geodesics and the ``shape'' of closed null orbits. To calculate these, we integrate the geodesic equations subjected to initial conditions,
\be
x_0^{\alpha}= \left(0, r_0,\pi/2, 0 \right) \,,\quad u_0^{\alpha}=\left(u_0^{\phi}/s, 0, 0, u_0^{\phi} \right)\,,
\ee
where $s=\sqrt{-g_{tt}(r_0, \pi/2,0)/g_{\phi \phi}(r_0, \pi/2,0)}$. These conditions ensure that we're integrating the motion for a null particle.
We then vary the affine parameter $\tau$ in an interval $ \left[0, \Upsilon \right]$ and fine-tune the value $r_0\sim 3M$ such that the orbit closes in the same interval. The period (in proper or affine time) $\Upsilon$ is roughly given by $\sim 2 \pi/ u_0^{\phi}$ and hence our calculation is independent from the initial value of $u^{\phi}$.

\begin{figure}[ht!]
\centering
\includegraphics[width=0.5\textwidth]{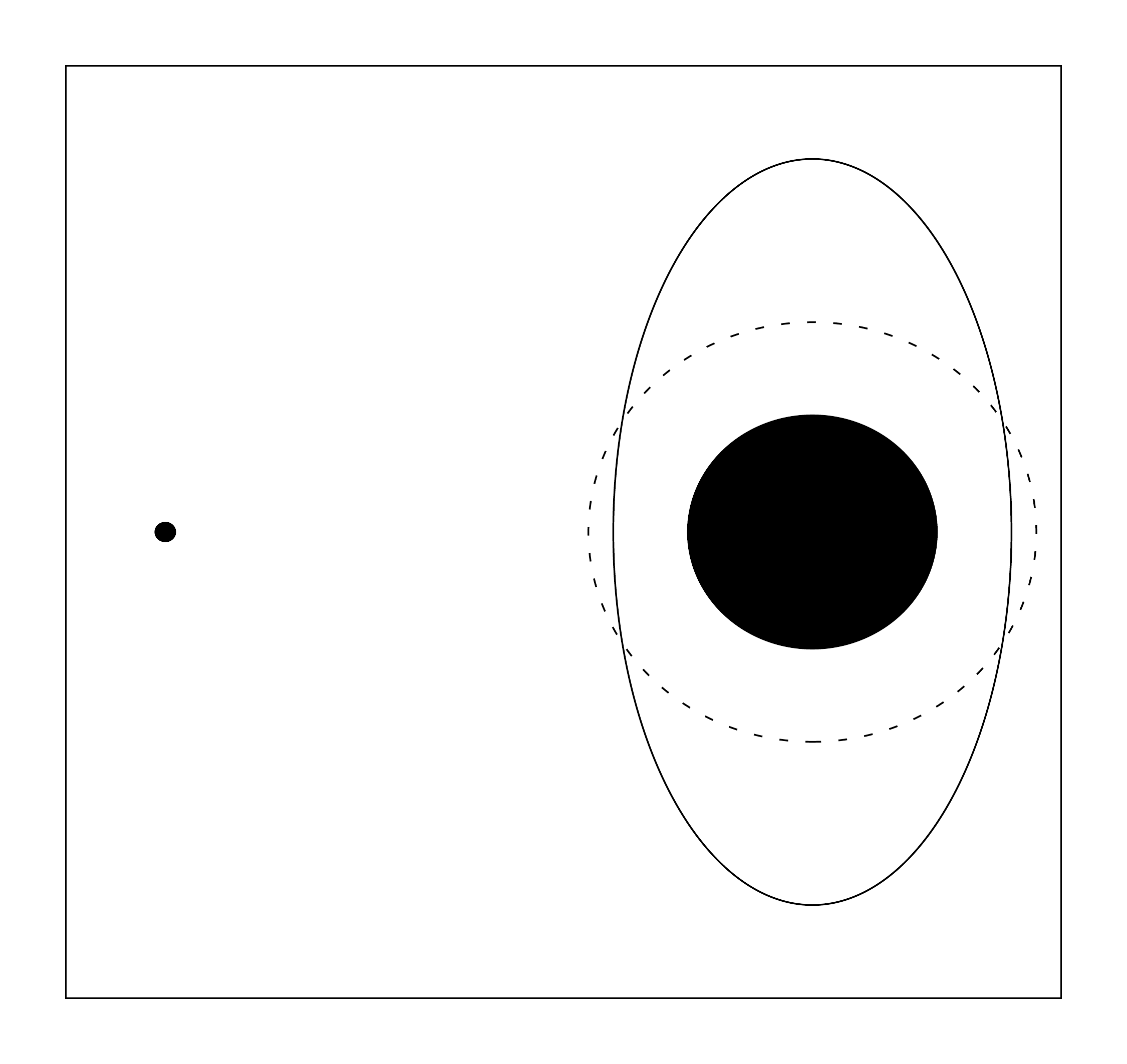}
\caption{Graphic representation of the closed null orbit \eqref{r_phi_true} (solid line) when the companion is fixed at the equator. The dashed circle represents the light ring in the Schwarzschild spacetime. Not in scale. \label{Fig:light_ring}}
\end{figure}
Our results show an interesting, perhaps expected feature (cf. Fig.~\ref{Fig:light_ring}): the null geodesic is no longer circular, but there {\it are} closed null orbits, which are elliptical. In fact, one can describe them analytically at small $\epsilon$, looking for an expansion with the functional form
\beq
r(\phi)&=& 3  M - \epsilon(x_0 + A \cos 2 \phi(\tau)) M\,,\label{r_phi}\\
\phi(\tau) &=& (c_1 + \epsilon c_2) \frac{\tau}{M} \, , \label{phi_tau} \\
t(\tau) &=& t_0 \tau  + \epsilon (t_1 + B \sin 2 \phi(\tau))M \, .\label{t_tau}
\eeq
The geodesic equations, together with the normalization for quadri-velocities, yield the solution
\beq
r(\phi) &=& 3  M - \epsilon \left(\frac{15}{2} + \frac{9}{2} \cos 2 \phi(\tau)\right) M \, , \label{r_phi_true} \\
t(\tau) &=& t_0 \tau  + \frac{45 \sqrt{3}}{4} \epsilon \sin 2 \phi(\tau) M \, ,\\
\phi(\tau) &=& t_0 \left(\frac{1}{3\sqrt{3}} - \frac{5}{6 \sqrt{3}} \epsilon \right) \frac{\tau}{M} \, ,
\eeq
where $t_0$ is a scale-factor of the affine parameter $\tau$ (with no influence on observables). Table~\ref{table_lightring} reports the values of $r_0$ (at $\phi=0$) obtained via numerical integration and the analytical prediction \eqref{r_phi_true}.
Once the solution (\ref{r_phi_true}) is known, one can find the parameters of the ellipse in terms of $\epsilon$. Namely, $a= (3- 3 \epsilon) M$ and $b=(3-12\epsilon) M$ are the semi-major and semi-minor axes respectively, while the eccentricity is given by $e =\sqrt{\frac{6 \epsilon}{1-4 \epsilon}}$.

Imposing $\phi(\Upsilon) = 2 \pi$, one can find the period of the orbit, which in the $t$-coordinate is
\beq 
T=t(\Upsilon) = 3 \sqrt{3} \pi (2+5 \epsilon) \, M \,.
\eeq
%

\subsection{Timelike geodesics: particles at rest}
%
\begin{table}[!htbp]
\caption{\label{table_static} Location $r_s$ where static equilibrium of particles in a tidally deformed BH spacetime is possible. Numerical values are denoted $r_s^{\rm num}$ and can be compared against the analytical prediction $r_s^{\rm an}$ of Eq.~\eqref{r_s}.}
\begin{ruledtabular}
\begin{tabular}{ccc}
\multicolumn{1}{c}{$\epsilon$} & \multicolumn{1}{c}{$r_s^{\rm num}/M$} & \multicolumn{1}{c}{$r_s^{an}/M$}                         
\\ \hline 
$10^{-6}$                      & 80.037                                & 79.713  \\
$10^{-5}$                      & 37.507                                & 37.195  \\ 
$10^{-4}$                      & 17.766                                & 17.479  \\  
$10^{-3}$                      & 8.604                                 & 8.372   \\
\end{tabular}
\end{ruledtabular}
\end{table}
The spacetime described by Eqs.~~\eqref{gtt_equat}-\eqref{gfifi_equat} seems to admit new types of motion. We find new results concerning static particles, i.e.,  orbits which satisfy $\dot{r}=\dot{\phi}=0$. At $\tau=0$, we set initial conditions $\phi= 0,\,r=r_s$ and the radial motion is governed by
\beq
\dot{r}^2 = E^2 + (2M - r_s) \left(\frac{1}{r_s} - \frac{2(r_s -2M) \epsilon}{M^2} \right)\, .
\eeq
Solving $\dot{r}=\ddot{r}=0$, one finds:
\beq
r_s &=& \left( \frac{2}{3}  + \frac{1}{(2 \epsilon)^{1/3}} \right)  M + \mathcal{O(\epsilon)} \, ,\label{r_s} \\ 
E^2 &=& 1 + \mathcal{O(\epsilon)}\, .
\eeq  

Estimates~\eqref{r_s} can be tested solving numerically the geodesic equations imposing $u_0^t = \sqrt{-1/g_{tt}^{\rm eq}(r_s, \pi/2,0)}$ as initial condition. Table~\ref{table_static} reports the results obtained with these two approaches, with an overall good agreement.

\begin{figure*}[ht!]
  \centering
  \includegraphics[width=0.45\textwidth]{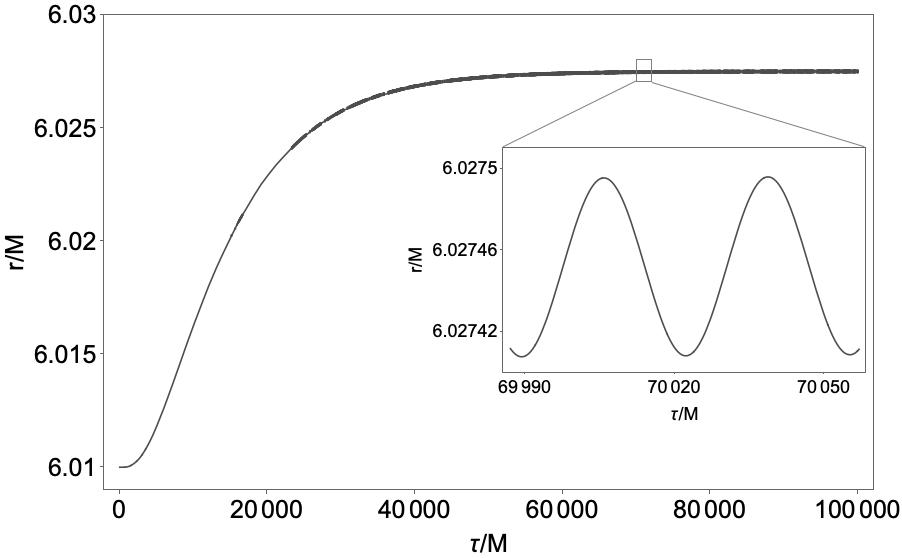}
  \includegraphics[width=0.45\textwidth]{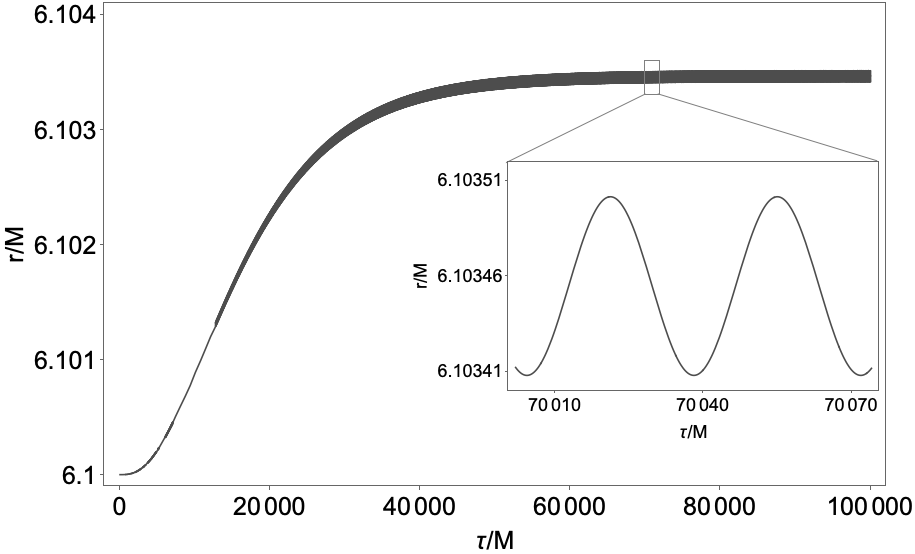}
\caption{Orbital radius of a particle, as function of proper time $\tau$. The particle is placed initially on a circular trajectory of radius $r_i=6.01 M$ (left panel) and $r_i=6.10 M$ (right panel)
and perturbed by a companion such that $R(\tau) = R_0/(1-e^{-\lambda(\tau + \delta)})$, with $\lambda = 10^{-4}$, $\delta = 10^{-4}$, $R_0 = 200 M$. The inset shows the behaviour of the radial coordinate on shorter timescales for $r_i/M = 6.01,  6.10$. Oscillations with periodicity $\sim 45M$ can be seen, which correspond to orbital periods $\sim 90M$ in the $(r, \phi)$-plane, in accordance with Fig.~\eqref{Fig:orbits} and Table~\ref{table_timelike}. Note that at early times there are some ``gaps'' in the oscillatory behaviour. These are a technical artifact of plot-drawing only and have no physical meaning.
}
\label{Fig:radius}
\end{figure*}
The result \eqref{r_s} is also in agreement with a Newtonian solution of the problem~\citep{butikov2002dynamical, hjorth1996elements}. Consider a binary system with masses $M$ and $M_c$ separated by a distance $R$. Let a test particle with mass $\mu$ be located at distance $r$ from the primary mass $M$. The test particle will be at rest if the gravitational force exerted by the primary mass
is compensated by the tidal force exerted by the secondary mass on the primary. 
Namely,
\beq
\frac{M \mu}{r^2} &=& \frac{M_c \mu}{(R-r)^2}-\frac{M_c \mu}{R^2}  \simeq \frac{2 M_c \mu r}{R^3} \, ,
\eeq
where we have exploited the fact that $r/R \ll 1$. Solving this equation for $r$ one finds:
\be
r=R \left(\frac{M}{2 M_c}\right)^{1/3} \,,
\ee
which is proportional to the value of $r_s$ in (\ref{r_s}) if we substitute in it the definition of $\epsilon$ given by \eqref{epsilon}.

\subsection{Timelike geodesics: disturbing circular motion}
%
\begin{table}[!htbp]
\caption{\label{table_timelike} Period $T$ in time coordinate and eccentricity-like parameter $e= \sqrt{1-(r_{\rm min}/r_{\rm max})^2}$ of the orbits performed by a timelike particle starting with initial radius $r_i$ when the companion is slowly approaching from infinity to $R_0=200M$.}
\begin{ruledtabular}
\begin{tabular}{ccc}
$r_i/M$  & 
 $T/M$ & $e$ \\ \hline 
 6.01 & 92 & 0.0055 \\
 6.05 & 95 & 0.0055 \\
6.1 
& 96 & 0.0057 \\
6.5 
& 104 & 0.0063 \\
7   
& 116 & 0.0074 \\
8   
& 142 & 0.0099 \\
10  
& 198 & 0.0148 \\
15  
& 363 & 0.0317 \\
20  
& 562 & 0.0522 \\
25  
& 786 & 0.0759 
\end{tabular}
\end{ruledtabular}
\end{table}
An analytical description of the ISCO is challenging to find in these non-symmetric setups. Without attempting to find an analytical solution to this problem, 
we wish to study when one disturbs circular orbits by slowly lowering a companion coming from infinity. This process could mimic for example the inspiral of a binary and its effect of the disk of one of them. 

We use a toy model of a time-dependent perturbation described by $\epsilon= M^2 M_c/R^3$ with $R$ a time-dependent quantity, 
\be
R(\tau)=\frac{R_0}{1-e^{- \lambda(\tau + \delta)}}\,.
\label{Rtau}
\ee
This behavior is meant to describe the appearance of a companion in a smooth way, so that one is able to study the transition from no-companion to a
tidally distorted BH. The results discussed below are not dependent on $\delta$, and converge to a universal behaviour for small $\lambda$. We focus now precisely on $(\delta, \,\lambda)$-independent results.

We set $\delta=M^2 \lambda = 10^{-4}M$, so that initially the companion is at $R\sim 10^8R_0$ and asymptotically approaches $R=R_0$.  
We focus on $R_0 = 200 M$.

\begin{figure*}[ht!]
 \centering
 \includegraphics[width=1\linewidth, valign=T, scale=1]{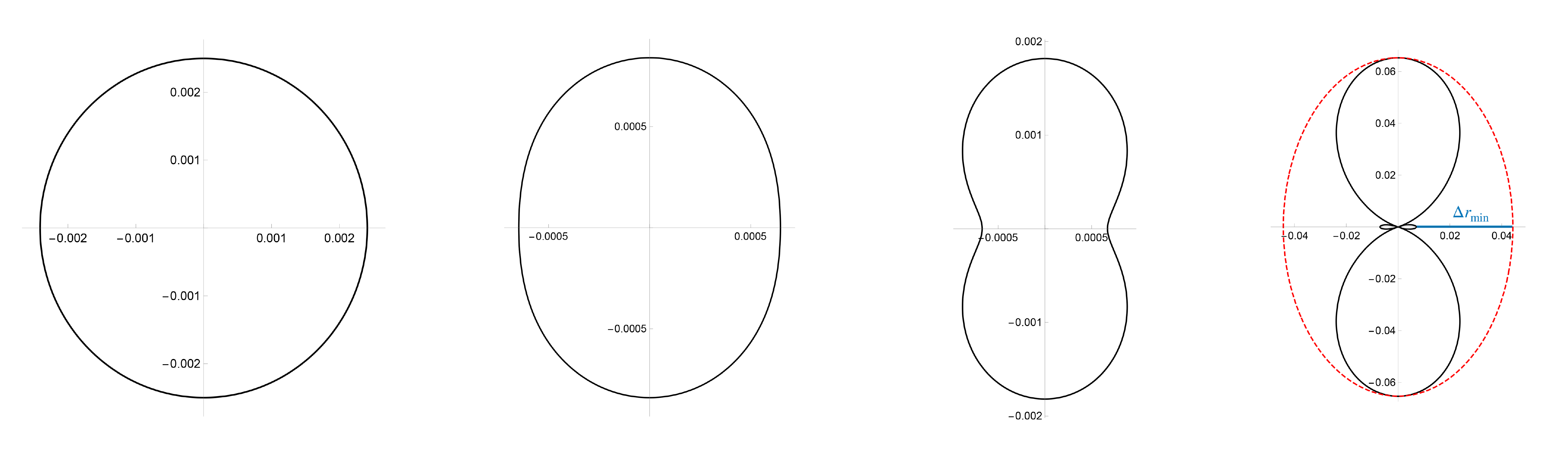}
\caption{Orbits in the $(r(\tau)-r_i, \phi(\tau))$ plane for initial radii (from left to right) $r_i/M =6.01, 7, 10, 25$ in a single period $T$. 
For graphical reason in the first case the radial coordinate is obtained as $r(\tau) - 6.025M$. In the last panel the blue line rapresents $\Delta r_{\rm min}$ mentioned in the main text.}
\label{Fig:orbits}
\end{figure*}
We start with a particle initially at $r=r_i$ on a circular orbit, i.e.,
\beq
u_0^{\mu} = \left( \sqrt{\frac{r_i}{r_i-3M}}, 0, 0, \frac{1}{r_i} \sqrt{\frac{M}{r_i-3M}} \right) \,,
\eeq
and we integrate the geodesic equation up to $\tau \sim 10 \lambda^{-1}$.

As the companion approaches its asymptotic location $R_0$, the particle is pulled to a slightly larger orbital radius, while the orbits becomes slightly eccentric, as shown in Fig.~\ref{Fig:radius}.

If we assume that the orbit is elliptic, we can extract the local maximum and minimum values of the radial coordinate to compute an eccentricity-like quantity%
\be
e = \sqrt{1-(r_{\rm min}/r_{\rm max})^2}\,,
\ee
as well as its period $T$. Specifically, these quantities are evaluated for a single period $T$ and extracted when the companion is almost steady, so at late times. Results are reported in Table~\ref{table_timelike}. Note that $T$ is to a good precision the orbital period of a particle in circular motion, given by Kepler's law.
Not surprisingly, $e$ increases as the particle moves away from the BH and get closer to the companion. Specifically, it scales as 
\beq
e \propto \left(\frac{r_i}{M} \right)^{1.8} \,.\label{ecc}
\eeq 

In fact, the motion is not really eccentric in these coordinates. For large $r_i$, we find that the motion acquires a peanut-like shape, as shown in Fig.~\ref{Fig:orbits}. 
If we take the $\phi = 0$ direction as the apoapsis $r_{\rm min}$ direction, then deviations from a perfect elliptic shape are of order $\Delta r_{\rm min}/r_{\rm min} \sim 10^{-3}$ (blue line in the last panel of Figure \ref{Fig:orbits}). As a consequence, the eccentricity-like parameter $e$ can still be considered an adequate measure of deviation from circularity.

Likewise, both the period $T$ and the shapes of the orbits in Figure \ref{Fig:orbits} are unaffected by the change in the final position. The only relevant difference is in the numerical values of $r_{\rm min}$ and $r_{\rm max}$, as well as in their ratio. In fact, as expected, these values become larger (smaller) if the companion is at $R_0/M = 100$ ($300$). As a consequence, orbits have a larger (smaller) value of the parameter $e$, though its scaling with $r_i$ is still well described by Eq.~\eqref{ecc}.  To be precise, the exponent is $1.71$ for $R_0 = 300$ and $1.88$ for $R_0 = 100$.

The final position of the particle, due to the presence of a companion, is dependent both on the strength of the perturbation and the initial radius. If we denote by $r_{\rm fin}$ the average value between $r_{\rm min}$ and $r_{\rm max}$, extracted when the companion is now steady, we find that for radii $r_i \geq 10 M$ the displacement $\Delta r =r_{\rm fin} - r_i$ is given by
\beq
\frac{\Delta r}{r_i} \approx A \, \left( \frac{r_i}{M} \right)^3 \, \epsilon
\eeq
with the constant $A$ of order of unity.

\section{Discussion}
The understanding of accretion disks or of the orbital motion of stars is a fundamental aspect in the interpretation of astrophysical observations. In addition, the location and properties of the 
the light ring around BHs are crucial in the understanding of the dynamics in the strong-field regime. Thus, peculiarities in motion in deep strong-field regions
could indicate either new physics or simply the presence of an unseen companion~\cite{Naoz:2019sjx}.
Motivated by this possibility, we studied the geodesic motion of both null and timelike orbits when tidal perturbations induced by an external companion are present.

There are several new features of tidally distorted BH geometries relative to BHs in isolation. 
When the orbital plane is orthogonal to that of the BH-companion, we were able to analytically quantify the linear deviations in the characteristic orbits induced by the companion. 
Specifically, with respect to the unperturbed configuration, the light ring has larger radius and orbital frequency, while the ISCO is located closer to the BH and, consequently, has a larger angular frequency.  In this framework we also obtained the QNM frequencies for the $l=m=2$ modes and we found that the ringdown frequency (the damping rate) is shifted to a slightly larger (smaller) value with respect to that of a Schwarzschild BH.

Instead, studying the late-time behaviour of the field showed that the structure of the modes of such a geometry is noticeably different from that of a single BH and is instead typical of confined systems. These properties had already been observed previously~\cite{Bernard:2019nkv}. However, in order to correctly capture all the physical properties of the relaxation stage, it is necessary to construct a spacetime which correctly reproduces the asymptotic behaviour at infinity~\cite{Bernard:2019nkv}. Further investigations in this direction are left for future work.

On the other hand, when orbits lie in the same plane as that of the BH-companion system, one needs a numerical study. We showed that the light ring is still a closed orbit but with an elliptical shape characterized by eccentricity and axes proportional to $\epsilon$ (cf. Fig.~\ref{Fig:light_ring}). Timelike orbits displayed a similar behaviour but other effects took place as well. The companion induces a relaxing of the orbit of point particles, which moves toward larger radii and are tidally distorted as seen in Fig.~\ref{Fig:orbits}.
We also explored the possibility for the spacetime to admit static orbits and we found an expression for their initial radius $r_s$ in agreement with the Newtonian treatment of the problem. 

\section*{Acknowledgements}
%
We acknowledge financial support provided under the European Union's H2020 ERC 
Consolidator Grant ``Matter and strong-field gravity: New frontiers in Einstein's 
theory'' grant agreement no. MaGRaTh--646597.
This project has received funding from the European Union's Horizon 2020 research and innovation programme under the Marie Sklodowska-Curie grant agreement No 101007855.
We thank FCT for financial support through Project~No.~UIDB/00099/2020.
We acknowledge financial support provided by FCT/Portugal through grants PTDC/MAT-APL/30043/2017 and PTDC/FIS-AST/7002/2020.
The authors would like to acknowledge networking support by the GWverse COST Action 
CA16104, ``Black holes, gravitational waves and fundamental physics.''
%
\appendix

\subsection{Geodesics in Majumdar-Papapetrou spacetimes\label{app:MP}}

There is a solution known in closed form, describing a regular and asymptotically flat BH binary spacetime geometry:
it is known as the Majumdar-Papapetrou geometry and describes a pair (or more) of charged, extremal BHs. The BHs feel no force
as they are extremal: their gravitational attraction is exactly canceled by an electrostatic repulsion.
In isotropic cylindrical coordinates the geometry reads
\beq
ds^2 = -\frac{1}{U^2} dt^2 + U^2 \left(d \rho^2 + \rho^2 d \phi^2 + dz^2 \right)\,,
\eeq
where
\be
U=1+\frac{m_1}{\sqrt{\rho^2 + (z + a)^2}}+ \frac{m_2}{\sqrt{\rho^2 + (z - a)^2}}\,.
\ee
Studies similar to that in the main body of this work were done in the context of such a spacetime~\cite{Assumpcao:2018bka,Nakashi:2019mvs,Nakashi:2019tbz}.
Following their analysis, we find the following corrections for the ISCO frequency of an extremal BH of mass $M$ perturbed by a companion of mass $M_c$ at a distance $R=2a$
\beq
M\Omega_{\rm ISCO}&=&\frac{\sqrt{3}}{16}\left(1-\frac{M_c}{2R}\right)\,,\\
M\Omega_{\rm LR}&=&\frac{1}{4}\left(1-\frac{M_c}{2R}\right)\,.
\eeq
These results have no approximation other than assuming a large separation $R$.
As one can see, the correction to the ISCO or light ring frequency has a scaling with $\epsilon$ (or equivalently, with distance $R$) different from that found in the main text for neutral binaries.
The disagreement can be traced back to the monopolar and dipolar components, and indirectly to the fact that this is not a purely gravitational system. 

To see this, perform a translation of the $z$ coordinate, $z \rightarrow z' - a$ and change from $(\rho, z')$ coordinates to the usual spheroidal coordinates $(r_{\rm iso}, \theta)$, to find
\be
U=1+\frac{m_1}{r_{\rm iso}} + \frac{m_2}{|\mathbf{r_{\rm iso}}- 2 \mathbf{a}|}\,.
\ee
Use the same Laplace expansion in~\eqref{laplace_potential} to expand the second term, but now neglect the monopole $l=0$ and the dipole $l=1$ terms, finding 
\be
U = 1 + \frac{m_1}{r_{\rm iso}} + \frac{m_2}{(2 a)^3} r_{\rm iso}^2 Y^*_{lm}(\theta_2, \phi_2) Y_{lm}(\theta, \phi)\,,
\ee
where again $(\theta_2, \phi_2)$ are the angular coordinates of the second mass $m_2$ in the new reference frame. We now want to express the metric in the new coordinates using the relations $\rho \rightarrow (r-m_1) \cos \theta$ and $z \rightarrow (r-m_1) \sin \theta$, where $r$ is the non-isotropic coordinate $r_{\rm iso} = r- m_1$. To make contact with the approach in the main text,
set $l=2$ in the harmonic expansion, fix the equatorial plane $\theta = \pi/2$ and specify the angular coordinates of $m_2$, which correspond to the polar case $(\theta_2, \phi_2) = (0,0)$. 
Assume large separations, $a \gg 1$, and expand the metric in powers of $(1/a)$ up to $\mathcal{O}(1/a^4)$. 
We then find the leading correction
\be 
m_1 \Omega_{\rm ISCO} = \frac{\sqrt{3}}{16} \left(1 + \frac{m_2}{R^3} \right)
\ee
which is now in perfect agreement (scaling-wise) with the main body, for neutral binaries. In conclusion, the Majumdar-Papapetrou spacetime does affect the geodesics in a different way, which can be ascribed to the spacetime not describing two BHs bound and evolving solely under the gravitational interaction. 

\bibliography{References}
\end{document}